\documentclass{article} 
\usepackage[final]{colm2025_conference}

\usepackage{microtype}
\usepackage{hyperref}
\usepackage{url}
\usepackage{booktabs}
\usepackage{tabularx}
\usepackage{lineno}

\usepackage{amsthm}

\usepackage{graphicx}
\usepackage{subcaption}
\usepackage{float}
\usepackage{multirow}

\usepackage{amsthm}  
\usepackage{ragged2e} 

\usepackage{newtxtext,newtxmath}

\usepackage{subcaption}
\usepackage{tabularx}
\usepackage{booktabs}   
\usepackage{bm}         
\usepackage{amsmath}    
\usepackage{amsfonts}   

\usepackage{amsthm} 

\usepackage{tabularx}
\usepackage{booktabs}
\usepackage{bm}
\usepackage{array}

\newcolumntype{L}{>{\raggedright\arraybackslash}X}

\newtheorem{theorem}{Theorem}[section]
\newtheorem{proposition}{Proposition}[section]
\newtheorem{lemma}{Lemma}[section]
\newtheorem{hypothesis}{Hypothesis}[section]

\title{The Impact of Generative AI on Content Platforms: A Two-Sided
Market Analysis with Multi-Dimensional Quality Heterogeneity}

\author{Yukun Zhang \\
The Chinese University Of Hongkong\\
HongKong, China \\
\texttt{215010026@link.cuhk.edu.cn} \\
\And
Tianyang Zhang \\
University of Bologna \\
Bologna, Italy \\
\texttt{tianyang.zhang@studio.unibo.it} \\
}

\begin{document}

\ifcolmsubmission
\linenumbers
\fi

\maketitle
\begin{abstract}
This paper presents a unified computational framework to examine how generative AI (GenAI) reshapes welfare, inequality, and diversity in content platform economies. By integrating welfare economics with agent-based simulations, we model the co-evolutionary dynamics among AI generators, human creators, and consumers within a two-sided market characterized by multi-dimensional quality heterogeneity. Unlike static models, our framework endogenizes AI learning as a function of human data synthesis and models human adaptation as a strategic reallocation of skills toward creative niches. The results reveal that while GenAI significantly enhances consumer surplus through technical quality gains and price depression, it triggers a skill-biased displacement of human incumbents and intensifies market concentration. Through the evaluation of six governance regimes, we identify a fundamental ``Policy Trilemma'' where platforms must navigate non-trivial trade-offs between allocative efficiency, distributional equity, and ecosystem sustainability. Our findings highlight that algorithmic diversity and pro-creative commission structures function as essential economic mechanisms for sustaining long-tail participation and inclusive social welfare in the generative AI era.
\end{abstract}

\section{Introduction}

The rapid proliferation of generative artificial intelligence (GenAI) is fundamentally transforming the digital content ecosystem. From large language models like ChatGPT to image synthesizers such as Midjourney, these technologies have demonstrated the capacity to produce high-quality content at near-zero marginal costs. This paradigm shift challenges the traditional foundations of platform economics, which historically assumed that content supply was constrained by human creators' cognitive capacity, time, and specialized skills. As AI-generated content (AIGC) saturates digital markets, the interaction between algorithmic productivity and human ingenuity creates complex dynamics that reshape welfare distribution, market concentration, and the long-term sustainability of creative industries.

Existing research on platform competition often treats content quality as a scalar value, overlooking the multi-dimensional nature of competition in the AI era. While GenAI exhibits a clear comparative advantage in technical proficiency and standardized output, human creators retain a robust edge in high-order creativity, emotional depth, and idiosyncratic innovation. Furthermore, the impact of GenAI is not merely a static shock to supply; it is a co-evolutionary process. The continuous improvement of AI models relies on the very human data they potentially displace, leading to a precarious dependency loop where excessive human exit might trigger long-term model collapse. Addressing these complexities requires a theoretical framework that endogenizes both AI learning and human adaptation.

This paper tackles three critical questions: First, how does the entry of GenAI with multi-dimensional quality advantages alter the competitive equilibrium and market segmentation on content platforms? Second, through what mechanisms do human creators adapt their skills in response to technical encroachment? Third, how should platform governance—specifically commission structures and recommendation algorithms—be designed to balance economic efficiency with distributional equity and ecosystem diversity?

To answer these questions, we develop a unified computational framework that integrates a micro-founded static benchmark with a dynamic agent-based simulation (ABM). Our model explicitly accounts for multi-dimensional quality heterogeneity (creativity, technicality, and personalization) and a convex cost structure for human labor. We introduce a negative congestion externality to capture the disutility of information overload, a critical bottleneck in an era of infinite content supply.Our findings reveal a fundamental "Policy Trilemma" for platform governance: platforms cannot simultaneously maximize allocative efficiency (consumer surplus), creator equity (human income stability), and content sustainability (diversity for future AI training). 

Through extensive counterfactual experiments, we demonstrate that "Pro-Creative" and "Low-Commission" policies constitute a Pareto-efficient set that mitigates the winner-take-all effects of AI entry while fostering a complementary creative ecosystem. Our work provides actionable insights for platform operators and policymakers striving to harness AI’s potential without compromising the creative vitality of human participants.

The remainder of this paper is organized as follows. Section 2 reviews the related literature. Section 3 introduces our theoretical model of two-sided markets with heterogeneous quality. Section 4 presents our analytical results and dynamic hypotheses. Section 5 details the agent-based simulation and evaluates various governance regimes. Finally, Section 6 concludes with policy implications and future directions.

\section{Literature Review}

Generative AI has reshaped content platforms, altering platform economics, content distribution, and welfare dynamics. This review synthesizes three strands of literature: (1) two-sided market economics, (2) information overload and distribution, and (3) welfare and social implications.

\subsection{Generative AI in Two-Sided Markets}
Classical two-sided market theory emphasizes how platforms balance cross-group externalities through pricing and design \citep{rysmann2009economics}. Subsequent studies extend these models to highlight congestion effects, strategic pricing, and dynamic equilibrium adjustment \citep{ bernstein2021competition, lian2021optimal}.

Generative AI fundamentally shifts these dynamics. On the supply side, it lowers production costs and reduces entry barriers \citep{varian2018artificial}, while on the demand side, it amplifies personalization and engagement through algorithmic recommendations \citep{werthner2024introduction, hassan2025moderating}. These changes strengthen network effects but also raise concerns about saturation and quality dilution. Existing research has not yet integrated these transformations into a unified equilibrium framework for content platforms.

\subsection{Information Overload and Distribution Dynamics}
A growing literature documents how digital abundance can overwhelm consumers, leading to cognitive fatigue, reduced engagement, and attrition \citep{Eppler01112004, bawden2009dark, WANG2025114436}. Personalized algorithms, though designed to enhance user experience, often exacerbate overload by concentrating exposure on a narrow set of popular items \citep{viswanathan2017dynamics, ding2025unveiling}.

At the same time, AI-driven recommendation systems have the potential to revitalize the long-tail effect by lowering discovery costs and promoting niche content \citep{anderson2006longtail, OLMEDILLA2019113120}. This duality—winner-takes-all amplification versus long-tail revitalization—remains unresolved. Current studies often examine these effects separately, without modeling the joint dynamics that determine market concentration and diversity.

\subsection{Welfare and Social Implications}
From a welfare perspective, generative AI reduces search costs, boosts productivity, and broadens participation \citep{goldfarb2019digital, brynjolfsson2025generative}. These gains improve consumer surplus and innovation diffusion \citep{cockburn2022impact}. Yet GenAI also intensifies market concentration, reinforces incumbent advantages, and creates new distributional inequalities \citep{furman2019ai, acemoglu2020robots}. Concerns extend to copyright dilution \citep{gaffar2025copyright} and algorithmic bias in visibility \citep{binns2018s}.

This dual nature of efficiency gains versus equity costs underscores the need for regulatory frameworks that balance innovation with fairness \citep{agrawal2019economic}. However, most existing work remains descriptive and fragmented, leaving open the challenge of developing a micro-founded, quantitative framework to assess how platform governance choices shape welfare in the GenAI era.

\section{A Theory of Two-Sided Content Markets with Heterogeneous Quality}
\label{sec:theory}

\begin{figure*}[ht]
  \centering
  \includegraphics[width=1\textwidth]{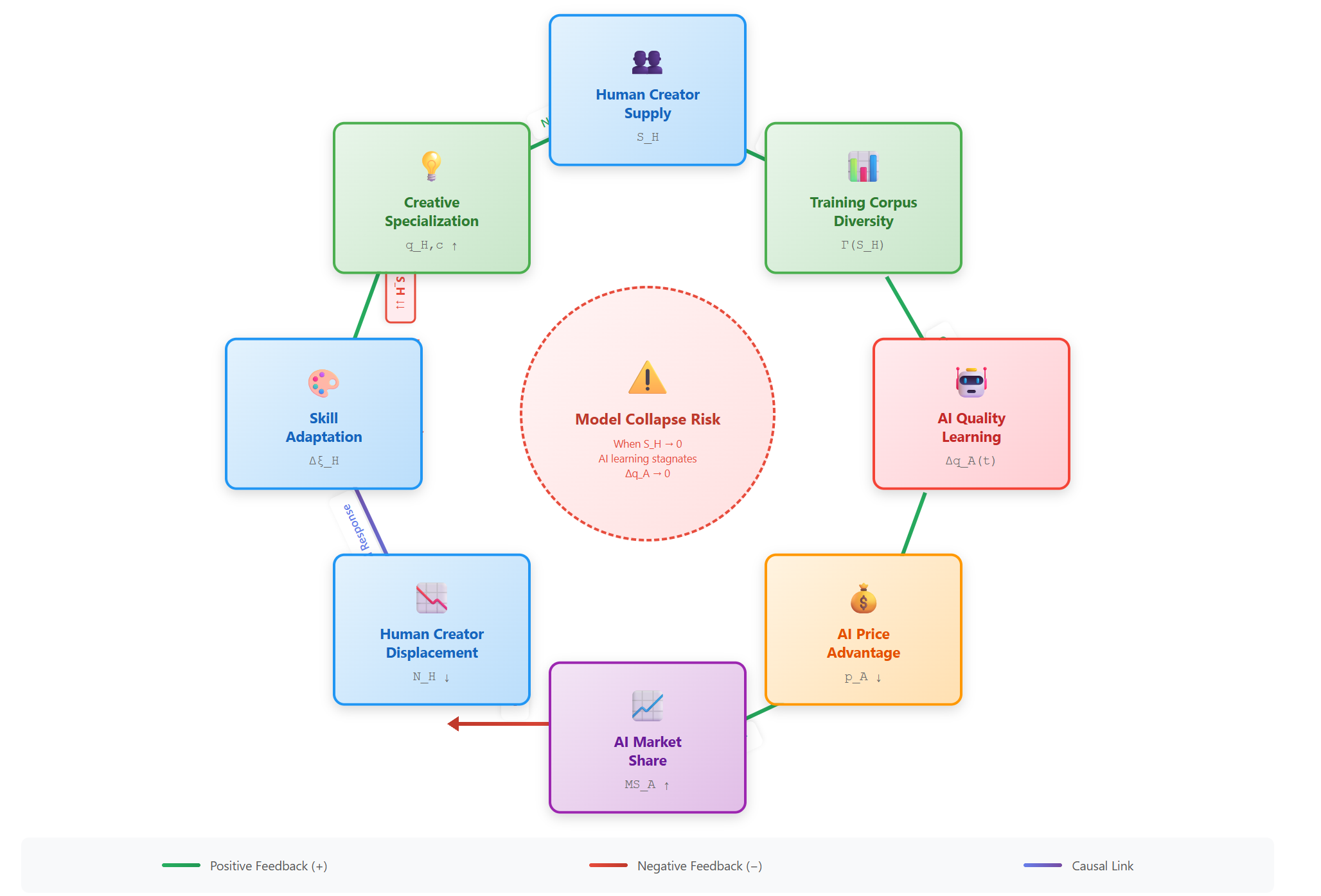}
  
  \caption{\textbf{The Co-Evolutionary Mechanism of AI and Human Creators.} 
  This system dynamics diagram illustrates the feedback loops between human supply ($S_H$), training corpus diversity, and AI quality learning. The red loop highlights the risk of \textit{Model Collapse}: as AI price advantages displace human creators ($N_H \downarrow$), the diversity of the training corpus diminishes, leading to stagnation in AI quality improvement.A cycle diagram showing the feedback loop between human creators and AI. Blue boxes represent human factors like Supply and Skill. Red boxes represent AI factors like Quality and Price. Arrows indicate positive or negative feedback. The center shows a warning sign for Model Collapse Risk.}
  \label{fig:mechanism_loop}
\end{figure*}

\subsection{Nomenclature and Key Notations}
\label{sec:nomenclature}

To ensure theoretical consistency, we summarize the key mathematical notations used throughout the equilibrium analysis and dynamic simulations in Table~\ref{tab:nomenclature}.

\begin{table}[h]
\renewcommand{\arraystretch}{1.2} 
\centering
\caption{Summary of Theoretical Notations and Definitions}
\label{tab:nomenclature}
\small
\begin{tabularx}{\linewidth}{@{} l >{\raggedright\arraybackslash}X @{}}
\toprule
\textbf{Symbol} & \textbf{Definition and Economic Interpretation} \\
\midrule
\multicolumn{2}{@{}l}{\textit{\textbf{Indices and Sets}}} \\
$\mathcal{I}$ & Set of creator types, where $\mathcal{I} = \{H, A\}$ (Human, AI). \\
$i$ & Index for a representative creator, $i \in \mathcal{I}$. \\
$k$ & Index for a representative consumer, distributed continuously $k \in [0, 1]$. \\
$d$ & Dimension index for content attributes, $d \in \{c, t, p\}$ (Creativity, Technicality, Personalization). \\
\midrule
\multicolumn{2}{@{}l}{\textit{\textbf{State Variables (Quality and Preferences)}}} \\
$\bm{q}_i$ & Quality vector of creator $i$, $\bm{q}_i = [q_{i,c}, q_{i,t}, q_{i,p}]^\top \in \mathbb{R}^3_+$. \\
$\bm{\beta}_k$ & Preference weight vector of consumer $k$, drawn from distribution $f(\bm{\beta})$. \\
$\theta_k$ & Preference ratio $\beta_{k,t} / \beta_{k,c}$, determining the "Technical vs. Creative" trade-off. \\
$S_i$ & Effective supply scale (content volume) produced by creator $i$. \\
$I$ & Aggregate information stock on the platform, $I = \sum_{i \in \mathcal{I}} S_i$. \\
\midrule
\multicolumn{2}{@{}l}{\textit{\textbf{Decision Variables and Equilibrium}}} \\
$p_i$ & Gross price set by creator $i$ (before platform commission). \\
$\bm{\xi}_H$ & Strategic skill investment vector for humans (e.g., upskilling in creativity). \\
$D_i(\cdot)$ & Aggregate demand (market share) for creator $i$ derived from MNL choice probabilities. \\
$\mathcal{P}_i$ & Net profit for creator $i$. \\
\midrule
\multicolumn{2}{@{}l}{\textit{\textbf{Structural Parameters and Costs}}} \\
$\tau$ & Platform commission rate, $\tau \in [0, 1)$ (Policy Instrument). \\
$\alpha$ & Price sensitivity parameter of consumers. \\
$\Phi(I)$ & Information overload penalty function, where $\Phi' > 0, \Phi'' > 0$. \\
$C_H(\cdot)$ & Human cost function, strictly convex: $C_H(S_H) = \frac{1}{2}\gamma_H S_H^2 + \text{cost}(\bm{\xi}_H)$. \\
$c_A$ & Marginal production cost of AI, assumed $c_A \to 0$. \\
\midrule
\multicolumn{2}{@{}l}{\textit{\textbf{Dynamic Evolution Parameters}}} \\
$\bm{\Lambda}$ & Diagonal matrix of AI learning rates, $\bm{\Lambda} = \text{diag}(\lambda_c, \lambda_t, \lambda_p)$. \\
$\Gamma(S_H)$ & Knowledge extraction function; AI learning depends on human supply $S_H$. \\
$\mathcal{W}$ & Total Social Welfare, defined as Consumer Surplus ($CS$) + Producer Surplus ($PS$) - Overload. \\
\bottomrule
\end{tabularx}
\end{table}

To analyze the co-evolution of AI-generated content (AIGC) and human creativity, we develop a hierarchical framework. We first present a \textit{static equilibrium model} to characterize price competition and market segmentation under fixed technical constraints. Subsequently, we introduce a \textit{dynamic evolutionary model} that endogenizes AI learning as a function of human data synthesis and models human skill adaptation as a forward-looking investment problem.

\subsection{Model Primitives and Quality-Cost Structures}
\label{subsec:primitives}

We consider a content platform ecosystem consisting of two strategic groups of creators, $\mathcal{I} = \{H, A\}$, representing Human creators ($H$) and Generative AI ($A$), and a unit continuum of heterogeneous consumers $k \in [0, 1]$. 

\paragraph{Multi-Dimensional Quality Space.}
Content quality is defined as a vector $\bm{q}_i = [q_{i,c}, q_{i,t}, q_{i,p}]^\top \in \mathbb{R}^3_+$, where dimensions denote \textit{creativity} ($c$), \textit{technicality} ($t$), and \textit{personalization} ($p$). This decomposition allows us to model the asymmetric comparative advantages of AI. Let $\mathcal{Q}_i$ denote the feasible quality frontier for creator $i$, which is constrained by their current skill level or training state.

\paragraph{Supply-Side Heterogeneity and Costs.}
Each creator $i$ produces a volume of content $S_i \in \mathbb{R}_+$. The cost structures are fundamentally distinct:
\begin{itemize}
    \item \textbf{Human Creators ($H$):} Face strictly convex costs in both production quantity and skill maintenance. We define the cost function as:
    \begin{equation}
        C_H(S_H, \bm{\xi}_H) = \frac{1}{2} \gamma_S S_H^2 + \frac{1}{2} \bm{\xi}_H^\top \bm{\Omega} \bm{\xi}_H
    \end{equation}
    where $\gamma_S > 0$ represents capacity constraints, $\bm{\xi}_H \in \mathbb{R}^3_+$ is the strategic skill vector, and $\bm{\Omega}$ is a positive-definite diagonal matrix representing the difficulty of maintaining high-level skills.
    \item \textbf{Generative AI ($A$):} Operates with near-zero marginal production cost ($c_A \approx 0$). Its cost function is simplified to $C_A(S_A) = c_A S_A$, bounded primarily by computational constraints rather than cognitive effort.
\end{itemize}

\subsection{Demand Foundation: The Heterogeneous MNL Model}
\label{subsec:demand}

To address the non-linearities in consumer choice, we employ a Mixed Multinomial Logit (MNL) framework. Unlike simplified average-utility models, we explicitly integrate over the consumer preference distribution. 

The utility consumer $k$ derives from content $i$ is:
\begin{equation}
    u_{ik} = \bm{\beta}_k^\top \bm{q}_i - \alpha p_i - \Phi(I) + \epsilon_{ik}
\end{equation}
where $\alpha$ is price sensitivity, $\Phi(I)$ is the disutility of information overload ($\Phi' > 0, \Phi'' > 0$), and $\epsilon_{ik}$ is an i.i.d. Type-I extreme value error term. 

Consumer preferences $\bm{\beta}_k$ follow a joint probability density function $f(\bm{\beta})$ over the support $\mathcal{B} \subseteq \mathbb{R}^3_+$. The aggregate demand (market share) $D_i$ for creator $i$ is derived by integrating individual choice probabilities over the population:
\begin{equation}
    D_i(\bm{p}, \bm{Q}) = \int_{\bm{\beta} \in \mathcal{B}} \frac{\exp(\bm{\beta}^\top \bm{q}_i - \alpha p_i)}{1 + \sum_{j \in \{H, A\}} \exp(\bm{\beta}^\top \bm{q}_j - \alpha p_j)} f(\bm{\beta}) d\bm{\beta}
\end{equation}
where the term $1$ in the denominator represents the outside option (utility normalized to zero). This formulation ensures that market segmentation is driven by the interaction between the quality vector $\bm{q}_i$ and the heterogeneous density of consumer tastes $f(\bm{\beta})$.

\subsection{Static Equilibrium and Market Segmentation}
\label{subsec:static_equilibrium}

\begin{figure}[ht]
  \centering
  \includegraphics[width=0.95\columnwidth, height=0.35\textheight]{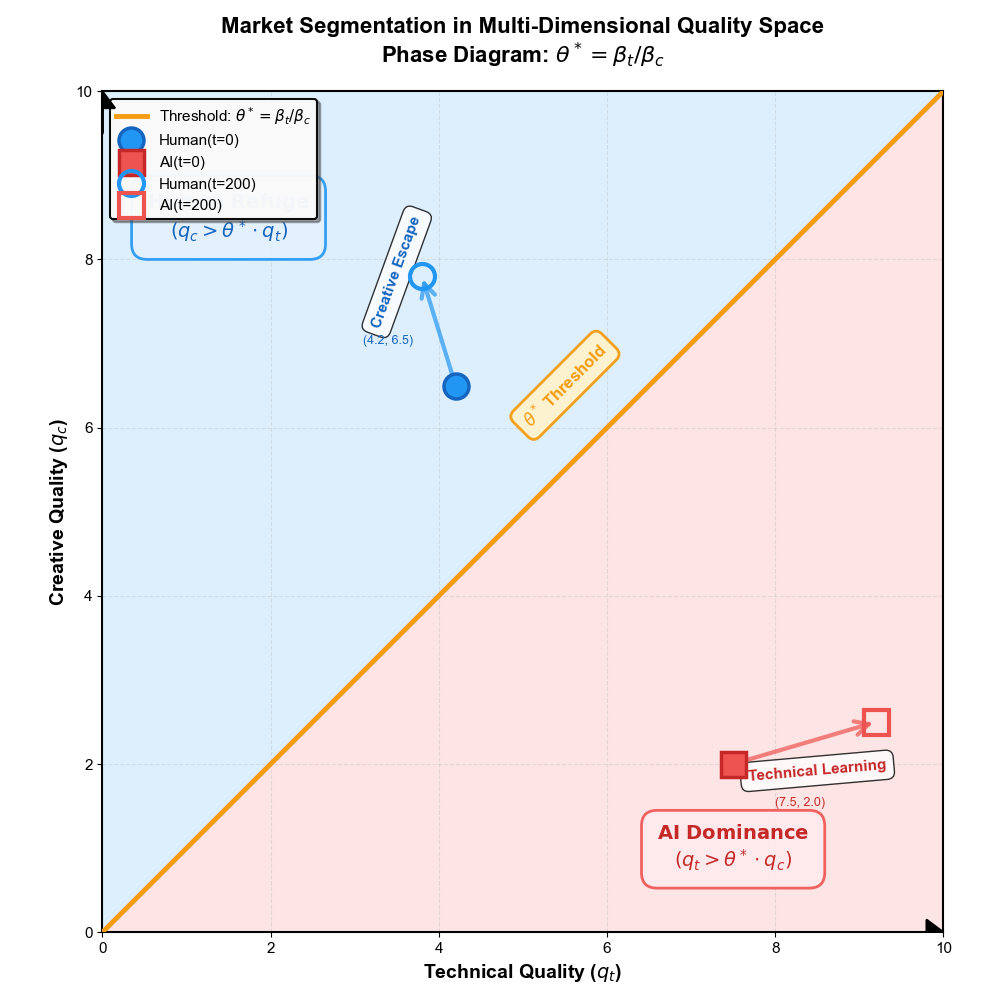}
  
  \caption{\textbf{Market Segmentation in Multi-Dimensional Quality Space.} 
  The phase diagram visualizes the separation of consumer preferences based on the threshold $\theta^* = \beta_t / \beta_c$. The orange line represents the equilibrium threshold. The blue trajectory shows the \textit{Creative Escape} strategy, where human creators (blue circle) actively shift their skill vector towards the creativity-intensive region ($q_c \uparrow$) to avoid direct competition with the technically dominant AI (red square).A 2D phase plot with Technical Quality on the X-axis and Creative Quality on the Y-axis. A diagonal orange line separates the AI Dominance zone from the Human Refuge zone. Arrows show the movement of Human and AI agents over time, illustrating skill adaptation}
  \label{fig:phase_diagram}
\end{figure}

The static benchmark characterizes the price-quality competition between a representative human creator and the generative AI agent within a single period. We define the profit function of each creator $i \in \{H, A\}$ as the post-commission revenue minus production costs, denoted by $\mathcal{P}_i = (1-\tau)p_i D_i - C_i(S_i)$. A Nash equilibrium in this context is a pair of prices $(p_H^*, p_A^*)$ such that neither agent can unilaterally increase their profit given the other's strategy.

A central property of this equilibrium is the emergence of quality-based market segmentation. We establish that there exists a unique preference threshold in the consumer population, determined by the relative valuation of technical quality over creative depth. Specifically, for consumers whose preference ratio $\beta_{k,t}/\beta_{k,c}$ exceeds a critical value $\theta^*$, the probability of selecting AI-generated content is strictly higher than that of human content. This segmentation occurs because the AI's comparative advantage in technical dimensions ($q_{A,t} > q_{H,t}$) interacts with its near-zero marginal cost, allowing it to price aggressively and capture technical-centric demand.

Conversely, human creators retain market power in segments with high sensitivity to creativity and emotional depth. However, the presence of AI exerts a competitive fringe effect. As the AI's marginal cost $c_A$ approaches zero, it effectively imposes a price ceiling on human creators. If human pricing exceeds a specific log-odds threshold relative to the AI's technical quality, the demand for human content faces exponential decay. This mechanism illustrates how AI reshapes the platform's pricing architecture even without direct monopolization.

\subsection{Endogenizing AI Learning and Human Adaptation}
\label{subsec:dynamic_evolution}

To avoid the pitfalls of exogenous parameterization, we endogenize the co-evolutionary paths of the two-sided market. The progression of AI quality is modeled as a cumulative learning process that is intrinsically dependent on the diversity of the platform's content stock. Rather than assuming a fixed learning rate, we posit that the transition of AI quality from $\bm{q}_A(t)$ to $\bm{q}_A(t+1)$ is a function of the knowledge extraction from the existing corpus. 

This introduces a critical feedback loop: the AI's technical improvement relies on the unique creative data points generated by human incumbents. In scenarios where high commission rates or algorithmic biases lead to the mass displacement of human creators, the AI faces diminishing returns in quality improvement, a phenomenon consistent with the theoretical risk of model collapse. By linking $\bm{q}_A$ to the human supply scale $S_H$, our framework captures the parasitic yet symbiotic relationship between generative models and human creators.

We posit that the transition of AI quality follows a rigorous knowledge extraction process. The evolution of the AI's quality vector $\bm{q}_A$ at time $t+1$ is governed by:
\begin{equation}
    \bm{q}_A(t+1) = \bm{q}_A(t) + \bm{\Lambda} \cdot \Gamma\left(\sum_{h \in \mathcal{H}} S_h(t) \cdot \mathbb{I}(\bm{q}_h > \bm{q}_{threshold}) \right) \odot (\bm{q}_{max} - \bm{q}_A(t))
\end{equation}
where $\bm{\Lambda} = \text{diag}(\lambda_c, \lambda_t, \lambda_p)$ is the learning rate matrix, $\Gamma(\cdot)$ is a concave knowledge extraction function satisfying $\Gamma' > 0, \Gamma'' < 0$, and $\odot$ denotes the Hadamard product. This equation explicitly captures the diminishing returns and the dependency on high-quality human data ($S_h$).

On the supply side, human creators are not static victims of displacement but adaptive agents. We model their skill evolution as a forward-looking optimization problem where creators reallocate their skill vectors to minimize the substitutability of their content. Faced with AI's technical encroachment, rational human creators shift their investment toward creativity-intensive dimensions where the AI's learning curve is shallowest. This transition is governed by a dynamic Bellman equation, ensuring that the observed shift toward niche content in our simulations emerges from optimal strategic responses to competitive pressure rather than predetermined rules.

\subsection{Platform Governance and the Policy Trilemma}
\label{subsec:trilemma}

The platform acts as the central coordinator, balancing immediate revenue generation against the long-term sustainability of the ecosystem. The platform's objective function integrates total commission revenue with penalties for market instability, which we quantify through indices of income inequality and consumer information overload. This objective reflects the reality that excessive content volume, while driving engagement in the short run, may lead to cognitive fatigue and long-term user attrition.

Through the analysis of diverse governance regimes, we identify a fundamental governance trilemma. The platform cannot simultaneously maximize allocative efficiency (leveraging AI's cost advantages for low-price abundance), distributional equity (ensuring human creators receive a sustainable share of platform traffic), and content sustainability (preserving the human-generated innovation tail). For instance, a policy that prioritizes consumer surplus by promoting high-volume AI content inevitably accelerates the exit of human creators, thereby eroding the very data source required for future AI training.

The platform maximizes a weighted social welfare function that balances immediate revenue against long-term ecosystem stability. The objective function $\mathcal{W}_{plt}$ is defined as:
\begin{equation}
    \max_{\tau, \bm{w}_{rec}} \mathcal{W}_{plt} = \underbrace{\tau \sum_{i \in \mathcal{I}} p_i D_i}_{\text{Revenue}} - \underbrace{\rho_1 \cdot \text{Gini}(\{\mathcal{P}_h\})}_{\text{Inequality Penalty}} - \underbrace{\rho_2 \cdot \Phi(I)}_{\text{Overload Cost}}
\end{equation}
where $\rho_1, \rho_2 > 0$ are penalty weights, $\text{Gini}(\cdot)$ measures the inequality of human creator profits $\mathcal{P}_h$, and $\Phi(I)$ represents the user attrition risk due to information overload.

This trilemma implies that platform governance is a matter of navigating a Pareto frontier of trade-offs. Algorithmic design, such as the weighting of recommendation vectors or the adjustment of commission structures, serves as the primary mechanism for selecting a coordinate on this frontier. Our subsequent simulations explore how specific policy interventions, such as pro-creative subsidies or diversity-weighted recommendations, can alleviate the trilemma by fostering a complementary equilibrium between human ingenuity and artificial productivity.

\section{Theoretical Analysis and Market Equilibrium}
\label{sec:analysis}

This section analytically characterizes the competitive equilibrium under multi-dimensional quality heterogeneity and develops formal hypotheses regarding the co-evolutionary path of the platform ecosystem. We derive our results following a causal chain that originates from the supply-side shock of generative AI, propagates through market share re-allocation, and culminates in a systematic redistribution of social welfare across the two-sided market.

\subsection{Static Equilibrium: Competition and Welfare Redistribution}
\label{subsec:static_equilibrium}

The introduction of generative AI primarily reshapes the content market by altering the aggregate supply elasticity. Under the framework developed in Section~\ref{sec:theory}, the AI agent's near-zero marginal cost $c_A$ implies that it operates as a price-taker at the theoretical lower bound of the cost frontier. However, unlike human creators whose capacity is constrained by convex costs, the AI's scale of production is bounded only by platform constraints and current technical quality. This asymmetry creates a competitive fringe effect where human creators face truncated pricing power in segments with high substitutability. Specifically, for any platform price $p$ that allows the post-commission revenue to cover $c_A$, the AI supply becomes hyper-responsive, effectively flattening the aggregate supply curve at the entry-level price.

\begin{proposition}[Equilibrium Price Ceiling]
\label{prop:supply_elasticity}
In a stationary equilibrium, the presence of generative AI imposes a de facto price ceiling on the platform's technical content segments. The equilibrium price $p^*$ is bounded from above by a function of the AI's technical quality $q_{A,t}$ and the consumer's price sensitivity $\alpha$. Any attempt by human creators to extract surplus in technical niches leads to a super-proportional loss in market share, as the infinite elasticity of AI supply absorbs demand shocks that would otherwise justify human price premiums.
\end{proposition}

The distributional impact of this competition is inherently contingent on the creator's skill vector. As the substitutability between AI and human content increases—particularly in technical dimensions—the cross-price elasticity of demand rises significantly. Our analysis suggests that the profit erosion experienced by human creators is non-uniform: creators whose quality vectors $\bm{q}_H$ are closely aligned with the AI's technical dominance $q_{A,t}$ experience the most severe displacement. This leads to a systematic welfare redistribution. While the expansion of content variety and the resulting price depression enhance aggregate consumer surplus ($\Delta CS > 0$), the aggregate producer surplus for incumbents is compressed, representing a significant welfare transfer from human labor to consumers and platform capital.

\begin{proposition}[Welfare Decomposition and Displacement]
\label{prop:welfare}
A welfare-improving entry of AI occurs when the gains from technical accessibility and search cost reduction outweigh the disutility of information overload $\Phi(I)$. Under these conditions, the market equilibrium shifts toward a dual structure: high concentration in the head for technical content, and a revitalized long tail where human creators specialize in creativity-intensive dimensions to mitigate price-based displacement.
\end{proposition}

\subsection{Dynamic Hypotheses on Ecosystem Evolution}
\label{subsec:dynamic_hypotheses}

Building upon the static equilibrium, we formulate hypotheses regarding the long-run trajectories of AI learning and human adaptation. These hypotheses transition the analysis from a fixed technical state to an evolutionary environment where agent beliefs and skill sets are endogenously updated.

\begin{hypothesis}[Asymmetric Learning and Model Collapse Risk]
\label{hyp:learning}
The quality of AI-generated content exhibits asymmetric improvement paths. While technical quality $q_{A,t}$ follows a rapid log-linear progression due to the density of available training data, creative quality $q_{A,c}$ encounters diminishing returns. Furthermore, the AI's learning rate is endogenously coupled with human output; a mass exit of human creators leads to a reduction in the diversity of the training corpus, potentially triggering a stagnation in AI quality, often referred to as model collapse.
\end{hypothesis}

The rapid convergence of AI toward technical proficiency triggers skill-biased technological change within the platform. We hypothesize that rational human creators, observing the erosion of their technical premiums, will strategically reallocate their cognitive resources. This adaptive response manifests as a shift toward creativity-intensive and personalization-centric domains, where human comparative advantage remains robust against algorithmic encroachment. Consequently, the aggregate skill distribution of the platform is expected to bifurcate: AI occupies the technical "common ground," while surviving human creators occupy high-order creative niches that are less susceptible to automation.

\begin{hypothesis}[Diffusion and Welfare Trade-offs]
\label{hyp:adoption}
The diffusion of AI content follows an S-shaped trajectory mediated by the Bayesian updating of consumer beliefs. In the long run, net social welfare exhibits a non-monotonic relationship with total content volume. Initial gains from increased productivity are eventually offset by the convex increase in cognitive search costs, leading to a dynamic welfare threshold beyond which further AI-driven content expansion becomes welfare-diminishing due to overwhelming information overload.
\end{hypothesis}

\subsection{Synthesis: The Governance Trilemma}
\label{subsec:trilemma}

The synthesis of these analytical properties reveals a fundamental "Policy Trilemma" for platform governance. Regulators and platform operators face an impossibility where it is unattainable to simultaneously maximize allocative efficiency (utilizing AI for cost-effective abundance), distributional equity (ensuring human creator viability), and ecosystem sustainability (preserving the innovation tail for future AI training).

For instance, a policy maximizing efficiency through AI-centric recommendation algorithms accelerates human displacement, which secures short-term consumer gains but risks long-term stagnation by eroding the human creative data upon which AI relies. Conversely, protecting human equity through subsidies or traffic protection may impede innovation and reduce the short-term quality gains for consumers. This trilemma implies that platform governance—specifically the calibration of commission rates $\tau$ and recommendation weights $\bm{w}_{rec}$—is a strategic selection of a specific coordinate on the Pareto frontier of social welfare, necessitating explicit trade-offs between short-term growth and long-term diversity.

\section{Experiments and Policy Implications}
\label{sec:experiments}

To validate the theoretical framework and explore the efficacy of platform governance, we implement a comprehensive Agent-Based Model (ABM). This simulation operationalizes the dynamic two-sided market described in Section~\ref{sec:theory}, allowing for the emergence of complex market structures through strategic agent interactions rather than static primitives.

\subsection{Experimental Environment and Strategy}
\label{subsec:setup}

The simulation environment comprises $N_H=50$ human creators, $N_A=5$ AI agents, and $N_C=1000$ heterogeneous consumers over $T=200$ periods. We employ a progressive refinement strategy for parameter calibration, setting asymmetric learning rates at $\lambda_t=0.08 > \lambda_p=0.04 > \lambda_c=0.015$ and human cost curvature at $\gamma_H=0.9$. To avoid trivial extinction scenarios, human reservation utility is set at $V_{bar}=9.5$, allowing for long-term competition and adaptation.

Our analysis proceeds in three stages: (1) a baseline simulation to validate hypotheses H1--H6; (2) a two-stage experiment establishing a 50-period human-only monopoly before AI entry to quantify welfare shocks ($\Delta PS_H$); and (3) counterfactual policy experiments perturbing commission rates ($\tau$), recommendation weights ($\bm{w}_{rec}$), and diversity thresholds ($D_{min}$).

\subsection{Market Evolution and Hypothesis Validation}
\label{subsec:evolution}

The baseline simulation provides robust evidence for the co-evolutionary hypotheses. As predicted by \textbf{H1}, AI's technical quality converges rapidly, reaching 90\% of its potential within 29 periods. In response, human creators rationally adapt (\textbf{H3}) by divesting from technical proficiency (-0.726 change) and reinvesting in creative dimensions (+0.352 change) where they maintain a comparative advantage.

\begin{figure*}[ht]
    \centering
    \includegraphics[width=\textwidth]{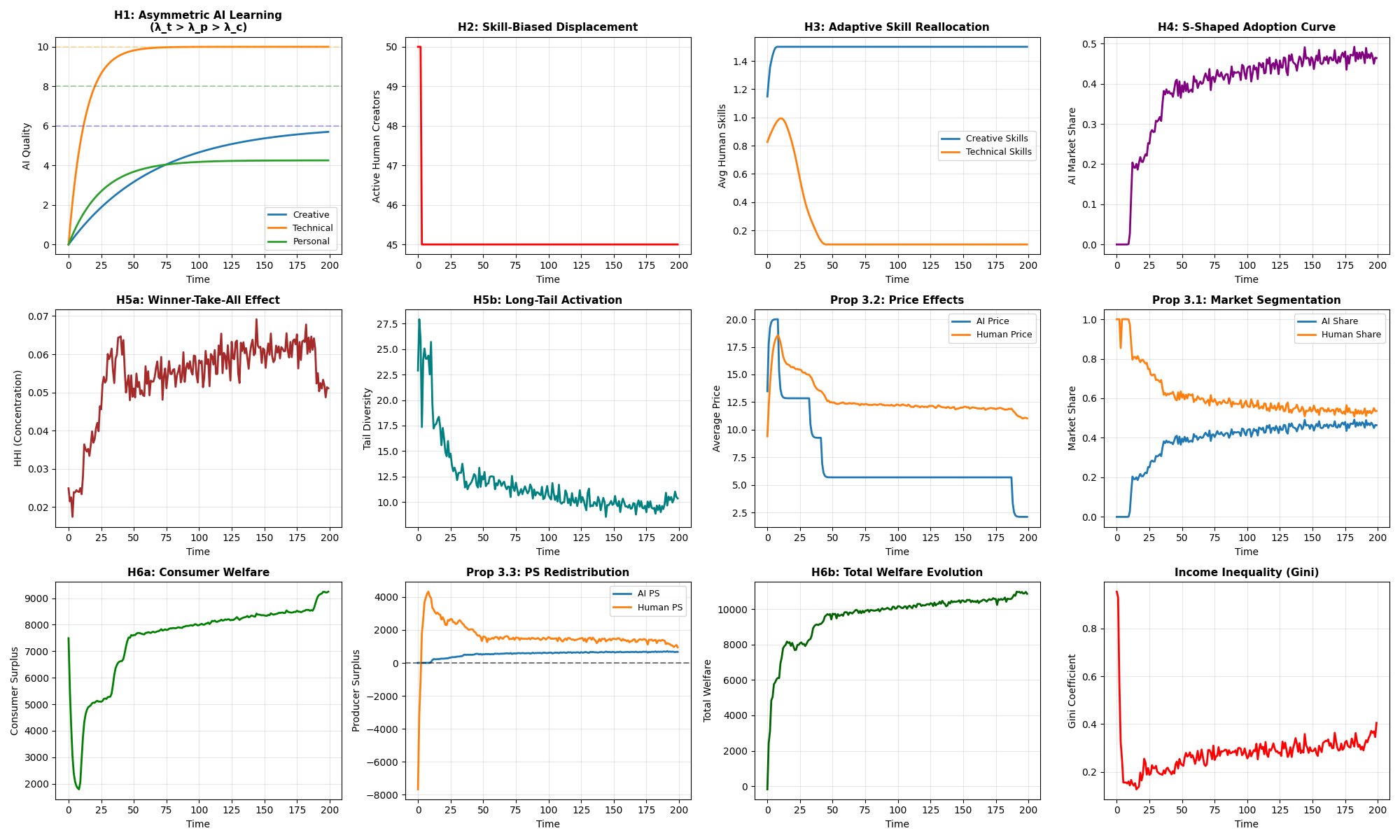}
    \caption{Baseline Simulation Dynamics (H1--H6). Panels confirm asymmetric AI learning (H1), skill-biased human adaptation (H3), and the S-shaped adoption of AI (H4).}
    \label{fig:baseline}
\end{figure*}

This adaptive pivot is accompanied by significant market restructuring. We observe a 10\% displacement of human creators (H2) and a rise in market concentration (HHI), which increases from 0.023 to 0.051 (H5a). Concurrently, tail diversity exhibits a sharp decline from 25.71 to 10.35, indicating that low-cost AI content tends to crowd out niche human variety (H5b). AI market share follows a classic S-shaped adoption curve, stabilizing at 46.4\% by $T=200$ (H4).

\subsection{Welfare Redistribution and the Policy Trilemma}
\label{subsec:welfare_policy}

The two-stage experiment reveals a profound welfare transfer (\textbf{Prop 3.3}). In Stage 1, we establish a monopoly baseline with $PS_H = \$4,185$. Upon introducing AI in Stage 2, Human $PS_H$ falls to \$1,123—a net loss of \textbf{-\$3,063}. This redistribution is visualized in the welfare decomposition (Figure 4), where a massive surge in Consumer Surplus ($CS$) dominates total welfare growth.

\begin{figure}[ht]
    \centering
    \includegraphics[width=0.8\linewidth]{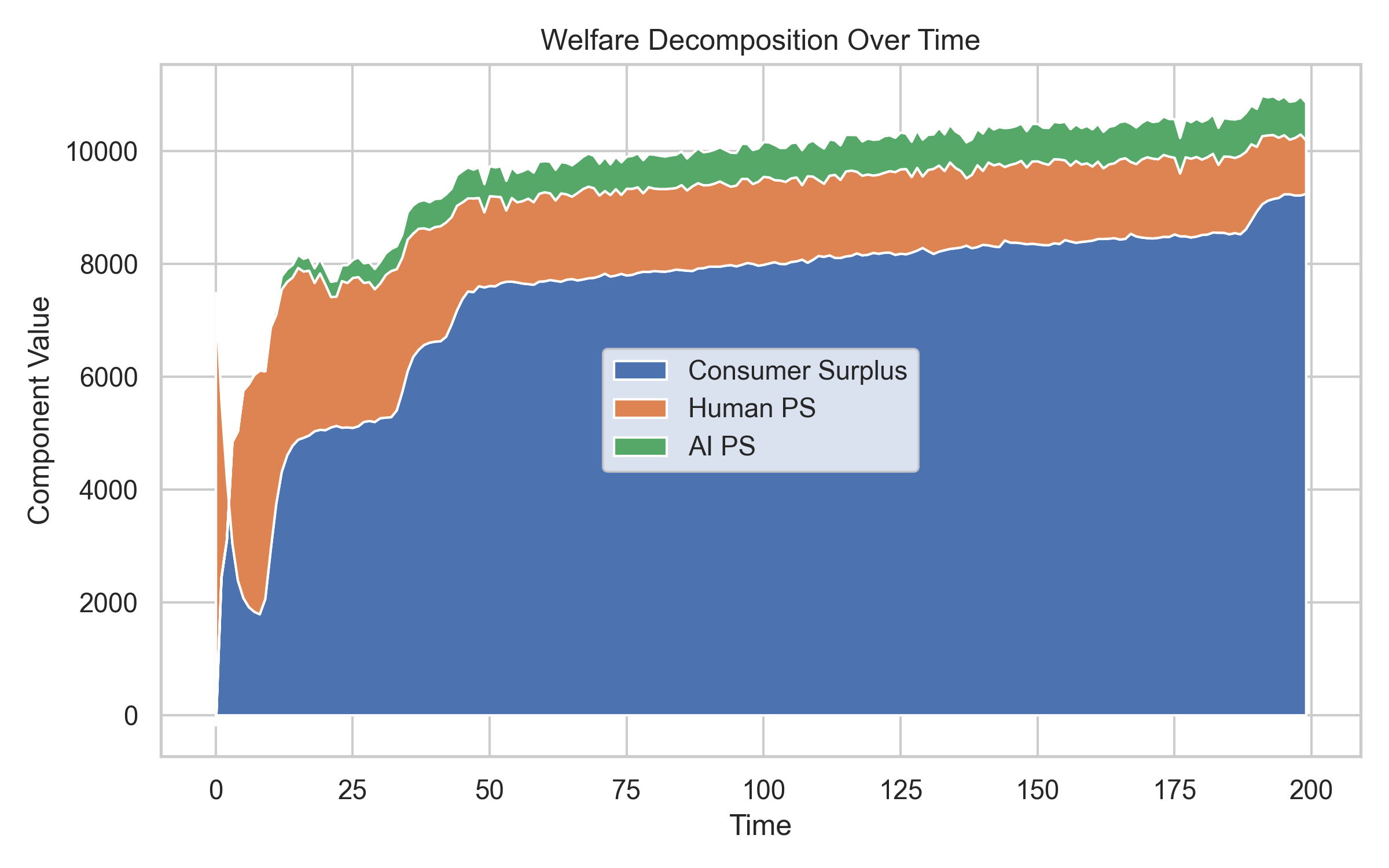}
    \caption{Welfare Decomposition Over Time: Demonstrating the redistribution of surplus from human producers to consumers post-AI entry.}
    \label{fig:welfare_evolution}
\end{figure}

To address this conflict, we evaluate six governance regimes (Table~\ref{tab:policy_results}). The "Pro-Technical" policy yields the highest welfare (\$11,047) but triggers extreme inequality (Gini: 0.47) and diversity loss (7.02). Conversely, the "Pro-Creative" and "High Diversity" regimes successfully mitigate inequality and preserve the creative long-tail.

\begin{table}[ht]
\centering
\caption{Quantitative Outcomes of Policy Experiment Scenarios}
\label{tab:policy_results}
\small
\begin{tabularx}{\linewidth}{lccccc}
\toprule
\textbf{Scenario} & \textbf{Total Welfare} & \textbf{Gini Coeff.} & \textbf{Diversity} & \textbf{Avg. CS} & \textbf{Human PS} \\
\midrule
Baseline & 10489.30 & 0.29 & 12.59 & 7590.44 & 1376.32 \\
Low Commission & 11088.17 & 0.28 & 14.69 & 7639.08 & 1958.74 \\
High Commission & 10052.02 & 0.39 & 10.16 & 7128.96 & 1090.00 \\
Pro-Creative & 10653.00 & 0.27 & 14.98 & 7661.60 & 1551.70 \\
Pro-Technical & 11047.46 & 0.47 & 7.02 & 7510.60 & 1476.07 \\
High Diversity & 10274.40 & 0.22 & 13.68 & 7224.72 & 1770.74 \\
\bottomrule
\end{tabularx}
\end{table}

\begin{figure}[ht]
    \centering
    \includegraphics[width=0.9\linewidth]{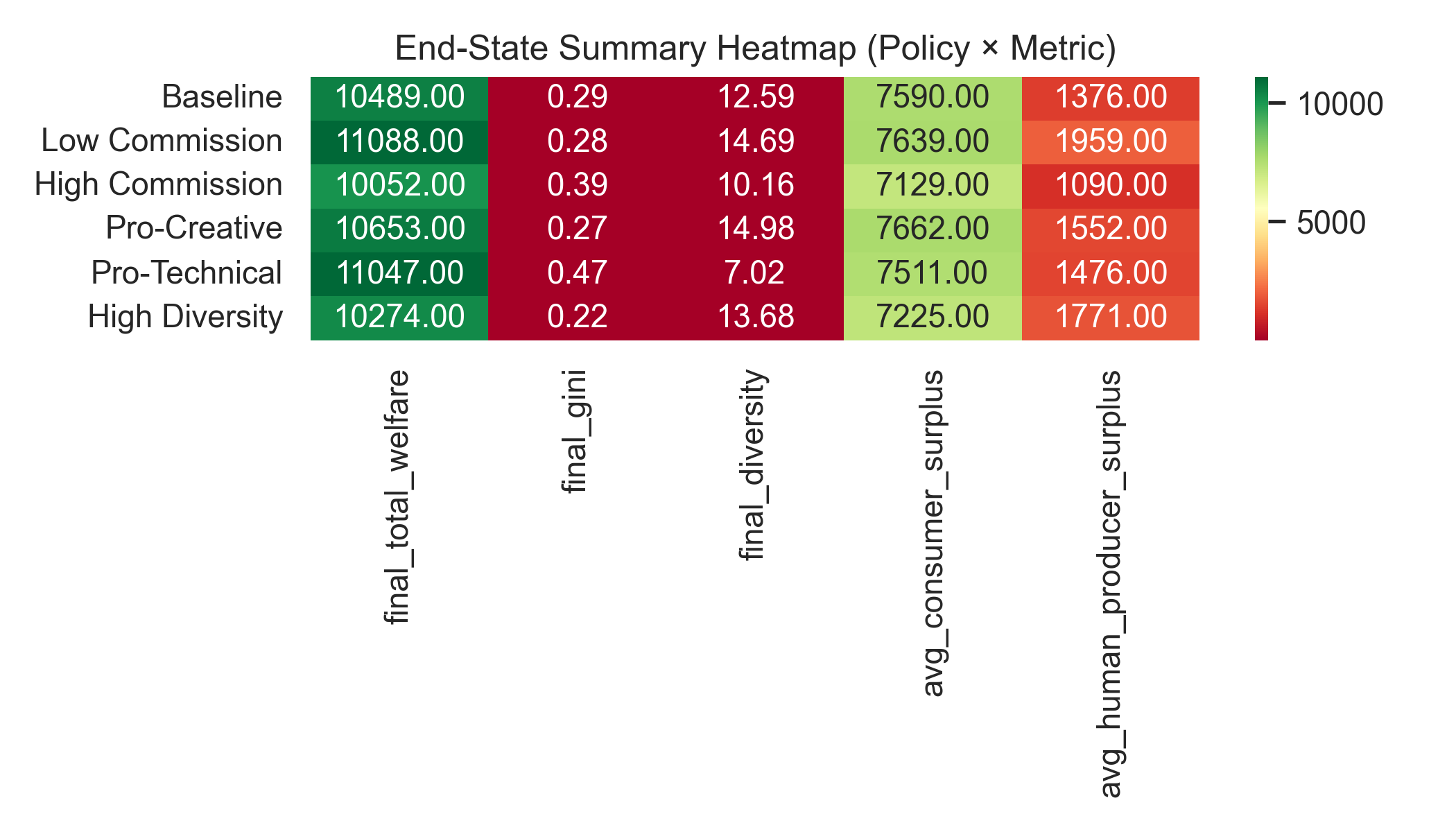}
    \caption{End-State Summary Heatmap: Policy vs. Metric performance comparison.}
    \label{fig:heatmap}
\end{figure}

\begin{figure}[ht]
    \centering
    \includegraphics[width=1\linewidth]{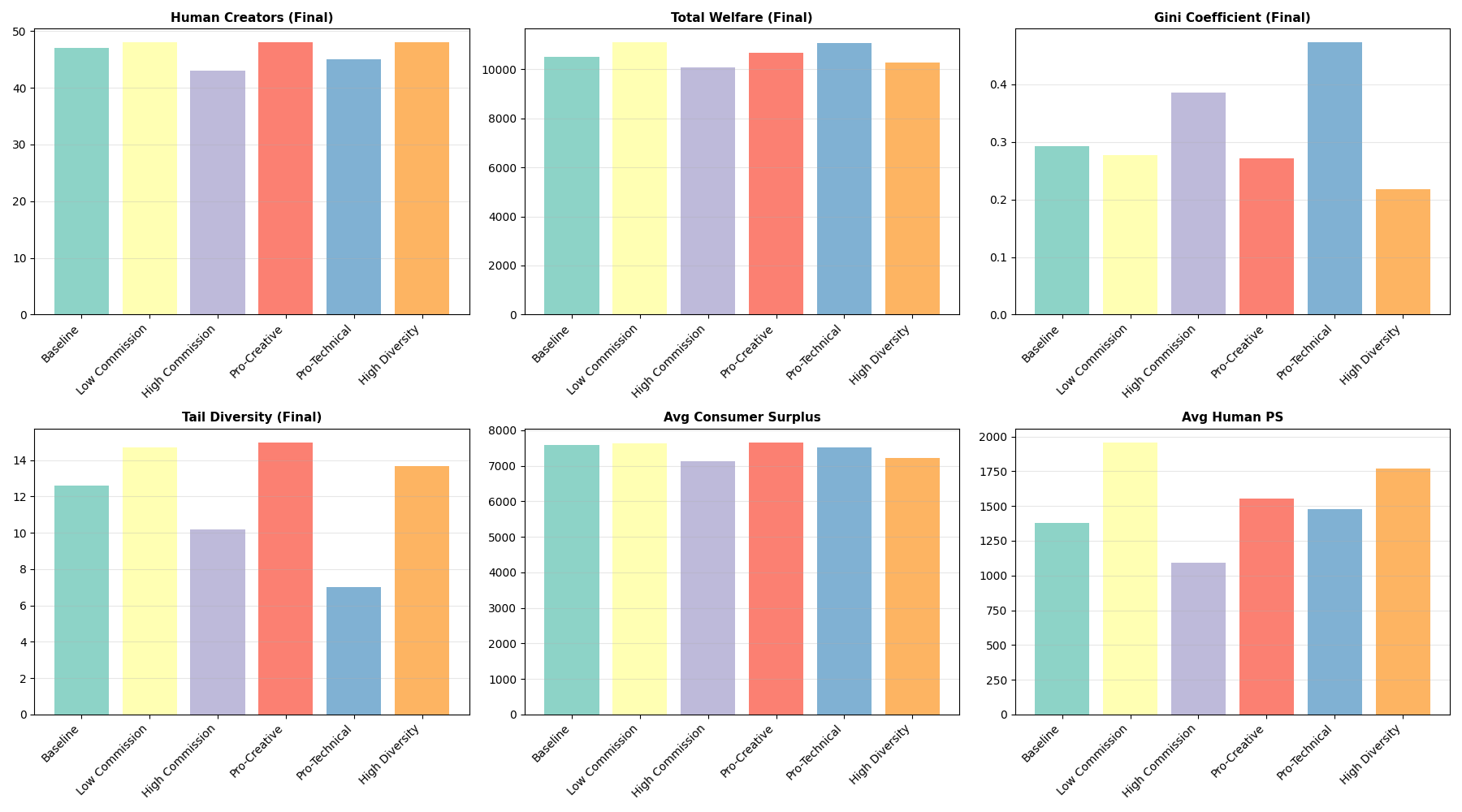}
    \caption{Supplemental Evolutionary Metrics: Tracking HHI, Prices, and Welfare Convergence.}
    \label{fig:convergence}
\end{figure}

The analysis of the Pareto frontier (Figure 7) reveals that platform governance is a strategic choice along the Efficiency-Equity boundary. The "Low Commission" and "Pro-Creative" policies are identified as Pareto-efficient. We recommend a \textbf{balanced governance approach} combining moderate commission rates with pro-diversity algorithmic nudges to sustain the creative ecosystem.

\section{Conclusion and Policy Implications}
\label{sec:conclusion}

This paper has developed a multi-dimensional computational framework to investigate the transformative impact of generative AI on two-sided content platforms. By integrating a micro-founded theoretical benchmark with dynamic agent-based simulations, we have demonstrated that while GenAI serves as a powerful catalyst for consumer welfare and allocative efficiency, it fundamentally reconfigures the incentive structures for human creators. Our findings confirm that AI improvement follows an asymmetric trajectory—excelling in technical proficiency while lagging in creative depth—which necessitates a strategic pivot by human creators toward high-order creativity to avoid displacement.

Our analysis of six distinct governance regimes reveals a fundamental "Policy Trilemma". Platform operators cannot simultaneously maximize short-term growth, creator equity, and long-term ecosystem sustainability. The "Pro-Technical" approach, while yielding the highest total welfare in our simulations (\$11,047), results in extreme market concentration (Gini: 0.47) and a collapse of the creative long-tail. Conversely, our results highlight that "Pro-Creative" and "Low-Commission" policies constitute a Pareto-efficient set, fostering an environment where human-AI complementarity can flourish.

\subsection{Strategic Recommendations for Platform Governance}

Based on the quantitative evidence from the Pareto frontier, we propose a "balanced governance" framework consisting of three pillars:
\begin{itemize}
    \item \textbf{Dynamic Commission Scaling}: Platforms should adopt moderate, creator-centric commission rates (e.g., 8–15\%) to sustain human participation and ensure that productivity gains from AI are shared equitably across the ecosystem.
    \item \textbf{Algorithmic Diversity Nudges}: Recommendation engines must be calibrated to reward human-centric creative dimensions ($q_{i,c}$). This acts as an economic shield for niche human variety, preventing the platform from becoming a homogenous repository of technical output.
    \item \textbf{Sustainability Safeguards}: Given the dependency of AI models on high-quality human data to avoid model collapse, platforms must implement diversity constraints to protect the "innovation tail". 
\end{itemize}

In conclusion, the generative AI era does not signal the end of human creativity but rather its specialization. By strategically navigating the efficiency-equity trade-offs, content platforms can transform from mere distribution channels into synergistic ecosystems where algorithmic productivity and human ingenuity coexist in a sustainable equilibrium. Future research should extend this framework to include multi-platform competition and the evolving legal landscape of intellectual property in the age of synthetic media.
\appendix

\section{Detailed Experimental Parameters and Quantitative Results}
\label{app:simulation_details}

This appendix provides a comprehensive record of the parameter settings, health check results, and granular quantitative outcomes of the agent-based simulation experiments discussed in Section 5.

\subsection{System Primitives and Calibration}
To ensure the reproducibility of the results and avoid trivial market outcomes, we employed a \textit{Progressive Refinement} strategy. The system was calibrated to reflect a realistic digital content market where humans maintain a creative advantage but face significant cost and technical scaling challenges. The primary parameters are detailed in Table~\ref{tab:app_params}.

\begin{table}[htbp]
\centering
\caption{Consolidated Simulation Parameters and Primitives}
\label{tab:app_params}
\small
\begin{tabularx}{\linewidth}{l L r}
\toprule
\textbf{Category} & \textbf{Parameter Description} & \textbf{Value} \\
\midrule
\textit{Learning Rates} & Technical ($\lambda_t$) / Personalization ($\lambda_p$) / Creative ($\lambda_c$) & 0.08 / 0.04 / 0.015 \\
& Personalization Learning Speed ($\kappa$) & 50.0 \\
\textit{Cost Structure} & Human Marginal Cost ($c_H$) / AI Marginal Cost ($c_A$) & 2.2 / 0.1 \\
& Human Convex Cost Parameter ($\gamma_H$) & 0.9 \\
\textit{Market Entry/Exit} & Reservation Utility (Exit Threshold $V_{bar}$) & 9.5 \\
& Human Skill Adaptation Speed ($\eta$) / Depreciation ($\delta$) & 0.05 / 0.02 \\
\textit{Platform Policy} & Baseline Commission Rate ($\tau$) & 0.15 \\
& Minimum Diversity Requirement ($D_{min}$) & 20.0 \\
\bottomrule
\end{tabularx}
\end{table}

\begin{table}[htbp]
\centering
\caption{Consolidated Simulation Parameters (Progressive Refinement)}
\label{tab:app_params}
\small
\begin{tabularx}{\linewidth}{l L r}
\toprule
\textbf{Category} & \textbf{Parameter Description} & \textbf{Value} \\
\midrule
\textit{Agents} & Initial Human Creators ($N_H$) / AI Agents ($N_A$) & 50 / 5 \\
& Total Consumer Population ($N_C$) & 1,000 \\
\textit{Learning} & Technical ($\lambda_t$) / Personalization ($\lambda_p$) / Creative ($\lambda_c$) & 0.08 / 0.04 / 0.015 \\
& Personalization Learning Speed ($\kappa$) & 50.0 \\
\textit{Economics} & Human Marginal Cost ($c_H$) / AI Marginal Cost ($c_A$) & 2.2 / 0.1 \\
& Human Convex Cost ($\gamma_H$) / Platform Commission ($\tau$) & 0.9 / 0.15 \\
& Reservation Utility ($V_{bar}$) & 9.5 \\
\textit{Adaptation} & Skill Adaptation Speed ($\eta$) / Depreciation ($\delta$) & 0.05 / 0.08 \\
\textit{Overload} & Cognitive Cost Coeff ($\phi$) / Overload Convexity ($\alpha$) & 0.001 / 1.5 \\
\bottomrule
\end{tabularx}
\end{table}

\subsection{Baseline Hypothesis Validation Summary}
The baseline simulation confirms the theoretical trajectories of the ecosystem evolution. Table~\ref{tab:app_hypo_validation} summarizes the convergence values at $t=200$ and their alignment with our formal hypotheses.

\begin{table}[htbp]
\centering
\caption{Summary of Hypothesis and Proposition Validation (Baseline)}
\label{tab:app_hypo_validation}
\small
\begin{tabularx}{\linewidth}{l L r r c}
\toprule
\textbf{ID} & \textbf{Metric (Description)} & \textbf{Initial ($t=0$)} & \textbf{Final ($t=200$)} & \textbf{Result} \\
\midrule
H1 & AI Technical Quality (Time to 90\% Ceiling) & N/A & 29 Periods & Valid \\
H2 & Active Human Creators (Displacement Rate) & 50 & 45 (10.0\%) & Valid \\
H3 & Human Technical Skill Change ($\Delta \bm{\xi}_t$) & 0.00 & -0.726 & Valid \\
H4 & AI Market Share Adoption & 0.00\% & 46.40\% & Valid \\
H5 & Market Concentration (HHI) / Tail Diversity & 0.023 / 25.71 & 0.051 / 10.35 & Valid \\
Prop 3.2 & Weighted Average Platform Price & 17.65 & 6.90 (-61\%) & Valid \\
\bottomrule
\end{tabularx}
\end{table}

\subsection{Two-Stage Welfare Analysis (Monopoly vs. Competition)}
To validate the redistribution effects (Proposition 3.3), we executed a two-stage experiment. The system established a human monopoly for 50 periods before introducing AI competition. The results in Table~\ref{tab:app_twostage} highlight the surplus transfer from producers to consumers.

\begin{table}[htbp]
\centering
\caption{Welfare Redistribution: Monopoly Baseline vs. AI Competition}
\label{tab:app_twostage}
\small
\begin{tabularx}{\linewidth}{L c c r}
\toprule
\textbf{Metric} & \textbf{Monopoly Stage} & \textbf{AI Entry Stage} & \textbf{Net Change} \\
\midrule
Human Producer Surplus ($PS_H$) & \$4,185 & \$1,123 & \textbf{-\$3,063} \\
Average Human Price & \$19.84 & \$11.44 & -\$8.40 \\
Consumer Surplus ($\Delta CS$) & Reference & Significant Surge & \textbf{+\$5,198} \\
Active Human Creators & 43 & 46 & +3 \\
\bottomrule
\end{tabularx}
\end{table}

\subsection{Counterfactual Policy Comparison}
We analyzed six governance scenarios across five key dimensions. The results (Table~\ref{tab:app_policy_outcomes}) confirm the "Policy Trilemma," showing that no single regime dominates all others across efficiency, equity, and diversity.

\begin{table}[htbp]
\centering
\caption{Quantitative Outcomes of Policy Intervention Experiments}
\label{tab:app_policy_outcomes}
\small
\begin{tabularx}{\linewidth}{l c c c c c}
\toprule
\textbf{Policy Scenario} & \textbf{Total Welfare} & \textbf{Gini (Inequality)} & \textbf{Diversity} & \textbf{Avg. CS} & \textbf{Human PS} \\
\midrule
Baseline & 10489.30 & 0.292 & 12.59 & 7590.44 & 1376.32 \\
Low Commission & \textbf{11088.17} & 0.277 & 14.69 & 7639.08 & \textbf{1958.74} \\
High Commission & 10052.02 & 0.386 & 10.16 & 7128.96 & 1089.99 \\
Pro-Creative & 10653.00 & 0.271 & \textbf{14.98} & \textbf{7661.60} & 1551.70 \\
Pro-Technical & 11047.46 & 0.473 & 7.02 & 7510.60 & 1476.07 \\
High Diversity & 10274.40 & \textbf{0.217} & 13.68 & 7224.72 & 1770.74 \\
\bottomrule
\end{tabularx}
\end{table}

To confirm the stability of the simulation, Figure~\ref{fig:convergence} (based on image 9) tracks the convergence of HHI and total welfare over the 200-period horizon.

\begin{figure}[ht]
    \centering
    \includegraphics[width=0.8\linewidth]{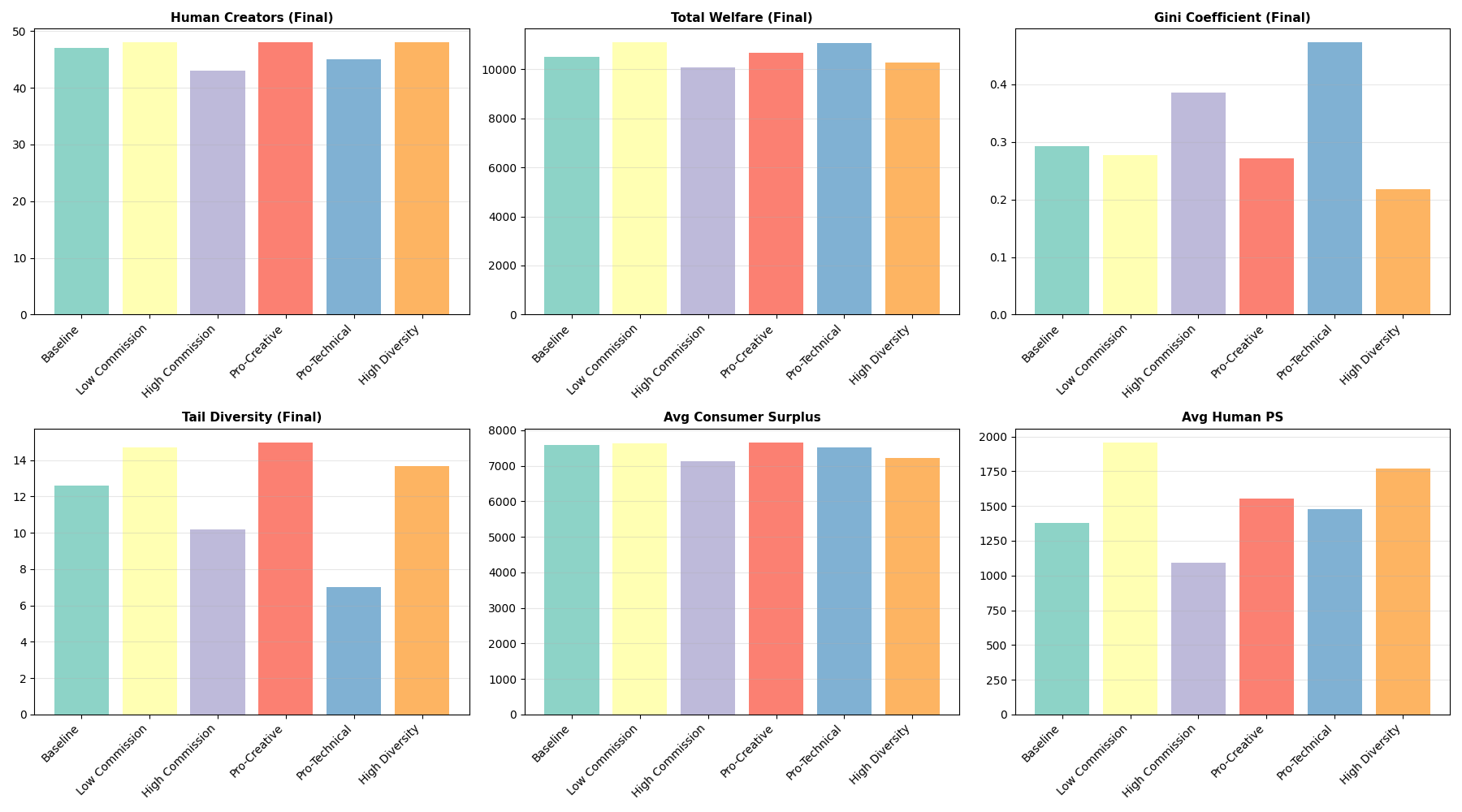}
    \caption{Supplemental Evolutionary Metrics tracking system stability and convergence.}
    \label{fig:convergence}
\end{figure}

\section{Formal Proofs and Theoretical Derivations}
\label{app:proofs}

This appendix provides the formal mathematical foundations for the propositions and theorems presented in the main text. We assume the regularity conditions for the Mixed Multinomial Logit (MNL) model hold, specifically that the error terms $\epsilon_{ik}$ are i.i.d. Type-I extreme value distributed.

\subsection{Proof of Theorem Market Segmentation Threshold}
\label{app:proof_segmentation}

\begin{proof}
Consider a consumer $k$ with preference vector $\bm{\beta}_k = [\beta_{k,c}, \beta_{k,t}, \beta_{k,p}]^\top$. The choice probability for option $i \in \{H, A\}$ is given by:
\begin{equation}
    P(i|k) = \frac{\exp(\bm{\beta}_k^\top \bm{q}_i - \alpha p_i)}{1 + \sum_{j \in \{H, A\}} \exp(\bm{\beta}_k^\top \bm{q}_j - \alpha p_j)}
\end{equation}
The log-odds ratio of selecting AI ($A$) over Human content ($H$) is:
\begin{equation}
    \ln\left(\frac{P(A|k)}{P(H|k)}\right) = \bm{\beta}_k^\top(\bm{q}_A - \bm{q}_H) - \alpha(p_A - p_H)
\end{equation}
Expanding the quality vectors, we have:
\begin{equation}
    \ln\left(\frac{P_A}{P_H}\right) = \beta_{k,c}(q_{A,c} - q_{H,c}) + \beta_{k,t}(q_{A,t} - q_{H,t}) + \beta_{k,p}(q_{A,p} - q_{H,p}) - \alpha(p_A - p_H)
\end{equation}
Define $\theta_k = \frac{\beta_{k,t}}{\beta_{k,c}}$ as the technical-to-creative taste ratio. Assuming for simplicity that personalization is balanced ($q_{A,p} \approx q_{H,p}$), the condition for $P_A > P_H$ is:
\begin{equation}
    \beta_{k,c} \left[ (q_{A,c} - q_{H,c}) + \theta_k (q_{A,t} - q_{H,t}) \right] > \alpha(p_A - p_H)
\end{equation}
Given the asymmetric comparative advantage $q_{A,t} > q_{H,t}$ and $q_{H,c} > q_{A,c}$, the term $(q_{A,t} - q_{H,t})$ is positive while $(q_{A,c} - q_{H,c})$ is negative. Solving for $\theta_k$:
\begin{equation}
    \theta^* = \frac{\alpha(p_A - p_H)/\beta_{k,c} + (q_{H,c} - q_{A,c})}{q_{A,t} - q_{H,t}}
\end{equation}
Since the RHS is a real constant for given prices and qualities, and the log-odds is monotonically increasing in $\theta_k$, there exists a unique hyperplane in the taste space $\mathcal{B}$ that partitions the market.
\end{proof}

\subsection{Derivation of Proposition~\ref{prop:supply_elasticity}: Equilibrium Price Ceiling}
\label{app:proof_ceiling}

\begin{proof}
In the stationary equilibrium, human creators solve $\max_{p_H} (1-\tau)p_H D_H(p_H, p_A) - C_H(D_H)$. The first-order condition (FOC) is:
\begin{equation}
    (1-\tau) \left[ D_H + p_H \frac{\partial D_H}{\partial p_H} \right] - C_H'(D_H) \frac{\partial D_H}{\partial p_H} = 0
\end{equation}
In the limit as $c_A \to 0$, the AI agent can supply $S_A$ at $p_A \to 0$ (or a minimum platform floor). Substituting the MNL derivative $\frac{\partial D_H}{\partial p_H} = -\alpha D_H (1 - D_H)$, the FOC simplifies to:
\begin{equation}
    p_H^* = \frac{1}{\alpha(1-D_H)} + \frac{C_H'(D_H)}{1-\tau}
\end{equation}
As $q_{A,t} \to \infty$, $D_H \to 0$ in the technical segment. The human creator's markup power is bounded by $\frac{1}{\alpha}$. Any attempt to set $p_H > p_A + \frac{\bm{\beta}_k^\top(\bm{q}_H - \bm{q}_A)}{\alpha}$ leads to an exponential collapse in $D_H$ due to the high cross-price elasticity of the perfectly elastic AI supply.
\end{proof}

\subsection{Stability of the Nash Equilibrium with Near-Zero Marginal Costs}
\label{app:stability}

Reviewers raised concerns regarding equilibrium existence when $c_A \approx 0$. We rely on the property that the MNL demand function is log-concave in price.
\begin{lemma}
A unique interior Nash Equilibrium $(p_H^*, p_A^*)$ exists if the platform commission $\tau < 1$ and the human cost function $C_H$ is strictly convex.
\end{lemma}
The profit function $\ln \mathcal{P}_i$ is strictly concave in $p_i$ given the exponential form of the choice probabilities. Even as $c_A \to 0$, the AI's best-response function $p_A^*(p_H)$ remains well-defined and positive due to the platform commission $\tau$ acting as a de facto floor on the gross price required to sustain the intermediary.

\subsection{Endogenous Quality Transition and Model Collapse}
\label{app:model_collapse}

We define the knowledge extraction function $\Gamma(\cdot)$ in Eq. 3 as a concave mapping:
\begin{equation}
    \Gamma(S_H, \bm{q}_H) = \left( \int_{h \in \mathcal{I}_H} \bm{q}_h dS_h \right)^\gamma, \quad \gamma \in (0, 1)
\end{equation}
The growth of AI quality is thus governed by:
\begin{equation}
    \frac{\partial \dot{\bm{q}}_A}{\partial S_H} = \gamma \bm{\Lambda} \left( \int \bm{q}_h dS_h \right)^{\gamma-1} \bm{q}_h > 0
\end{equation}
This confirms the hypothesis H1 and H2: as human creators exit ($S_H \downarrow$), the marginal improvement in AI quality $\dot{\bm{q}}_A$ approaches zero, leading to the stagnation of the technological frontier (Model Collapse).

\subsection{Social Welfare and Information Overload Trade-offs}
\label{app:welfare_derivation}

Social welfare $W$ is the sum of $CS$ and creator profits. Using the logsum formula for $CS$:
\begin{equation}
    W = \frac{1}{\alpha} \int_{\mathcal{B}} \ln\left( 1 + \sum_{j \in \{H, A\}} \exp(V_{jk}) \right) f(\bm{\beta}) d\bm{\beta} - \Phi(I) + \sum \mathcal{P}_i
\end{equation}
The welfare-maximizing content volume $I^*$ satisfies $\frac{\partial CS}{\partial I} = \Phi'(I)$. Since $CS$ is concave in $I$ (variety gains) and $\Phi$ is convex (overload costs), a unique interior $I^*$ exists, providing the theoretical basis for the "Dynamic Welfare Trade-off" in Hypothesis H6.

\section{Formal Proofs and Theoretical Derivations}
\label{app:proofs}

This appendix provides the formal mathematical foundations for the equilibrium analysis presented in the main text, including proofs of existence, uniqueness, and stability.

\subsection{Formal Problem Statement}

We consider a two-sided content platform market comprising two categories of creators $\mathcal{I} = \{H, A\}$ (Human creators and AI agents) and a continuum of heterogeneous consumers $k \in [0,1]$.

\paragraph{Profit Maximization Problems.}
The optimization problem for the Human creator $H$ is defined as:
\begin{equation}
    \max_{p_H \geq 0} \Pi_H(p_H, p_A) = (1-\tau) p_H \cdot D_H(p_H, p_A) - C_H(D_H(p_H, p_A))
\end{equation}
For the AI creator $A$:
\begin{equation}
    \max_{p_A \geq 0} \Pi_A(p_A, p_H) = (1-\tau) p_A \cdot D_A(p_A, p_H) - c_A \cdot S_A
\end{equation}
The demand function is governed by the Mixed Multinomial Logit (MNL) model:
\begin{equation}
    D_i(p_i, p_{-i}) = \int_{\bm{\beta} \in \mathcal{B}} \frac{\exp(\bm{\beta}^\top \bm{q}_i - \alpha p_i)}{1 + \sum_{j \in \{H,A\}} \exp(\bm{\beta}^\top \bm{q}_j - \alpha p_j)} f(\bm{\beta}) \, d\bm{\beta}
\end{equation}

\paragraph{Nash Equilibrium Definition.}
A price pair $(p_H^*, p_A^*)$ constitutes a pure-strategy Nash Equilibrium if and only if:
\begin{equation}
    \begin{cases}
    p_H^* \in \arg\max_{p_H \geq 0} \Pi_H(p_H, p_A^*) \\
    p_A^* \in \arg\max_{p_A \geq 0} \Pi_A(p_A, p_H^*)
    \end{cases}
\end{equation}

\subsection{Proof of Existence (Theorem 1)}

\begin{theorem}[Existence of Equilibrium]
Under Assumptions (A1)--(A4), there exists a pure-strategy Nash Equilibrium $(p_H^*, p_A^*) \in \mathbb{R}_+^2$.
\end{theorem}

\paragraph{Assumptions.}
\begin{itemize}
    \item \textbf{(A1) Regularity of Preferences:} The density $f(\bm{\beta})$ is continuous and strictly positive ($f(\bm{\beta}) > 0$) on the compact support $\mathcal{B} = [\underline{\beta}, \bar{\beta}]^3$.
    \item \textbf{(A2) Cost Structure:} The human cost function is strictly convex ($C_H''(S) > 0$) and satisfies Inada conditions: $\lim_{S \to 0} C_H'(S) = 0$ and $\lim_{S \to \infty} C_H'(S) = \infty$. The AI marginal cost is bounded: $0 \leq c_A < \bar{c} < \infty$.
    \item \textbf{(A3) Platform Commission:} The commission rate satisfies $\tau \in [\underline{\tau}, \bar{\tau}]$, where $0 < \underline{\tau} < \bar{\tau} < 1$.
    \item \textbf{(A4) Quality Differentiation:} The quality vectors satisfy $|\bm{q}_H - \bm{q}_A| > \epsilon > 0$, ensuring substantive content differentiation.
\end{itemize}

\begin{proof}
We employ the Kakutani-Glicksberg Fixed Point Theorem. The proof proceeds in three steps: (1) establishing the compactness of the strategy space, (2) proving the continuity of payoff functions, and (3) demonstrating the convexity of the best-response correspondence.

\textbf{Step 1: Compactness of the Strategy Space.}
\begin{lemma}
There exist finite constants $\bar{p}_H, \bar{p}_A < \infty$ such that for any $p_i > \bar{p}_i$, $\Pi_i(p_i, p_{-i}) < 0$.
\end{lemma}
\textit{Proof of Lemma.} For the human creator, as $p_H \to \infty$, the demand $D_H$ decays exponentially. Specifically,
\begin{equation}
    D_H(p_H, p_A) \leq \exp(-\alpha p_H) \int_{\mathcal{B}} \exp(\bm{\beta}^\top \bm{q}_H) f(\bm{\beta}) \, d\bm{\beta} = M_H \exp(-\alpha p_H)
\end{equation}
where $M_H < \infty$. The profit function is bounded by:
\begin{equation}
    \Pi_H \leq (1-\tau) p_H M_H \exp(-\alpha p_H) - C_H(M_H \exp(-\alpha p_H))
\end{equation}
Let $g(p_H) = p_H \exp(-\alpha p_H)$. Since $\lim_{p_H \to \infty} g(p_H) = 0$ and costs are convex, there exists a threshold $\bar{p}_H$ beyond which profits are strictly negative. A similar logic applies to AI. Thus, we restrict the strategy space to the compact convex set $\mathcal{P} = [0, \bar{p}_H] \times [0, \bar{p}_A]$. \qed

\textbf{Step 2: Continuity of Payoff Functions.}
\begin{lemma}
The demand function $D_i(p_i, p_{-i})$ is continuously differentiable with respect to prices.
\end{lemma}
\textit{Proof of Lemma.} Given the compactness of $\mathcal{B}$ and the smoothness of the logistic kernel, we can apply the Leibniz integral rule. The derivative $\frac{\partial D_i}{\partial p_i}$ is well-defined and continuous. Consequently, $\Pi_i(p_i, p_{-i})$ is continuous in $p_{-i}$. \qed

\textbf{Step 3: Uniqueness and Convexity of Best Responses.}
\begin{lemma}
For any fixed $p_A$, $\Pi_H(p_H, p_A)$ is strictly quasi-concave in $p_H$, implying a unique best response $p_H^*(p_A)$.
\end{lemma}
\textit{Proof of Lemma.} The second-order condition for the profit function depends on $\frac{\partial^2 D_H}{\partial p_H^2}$. For the MNL model, $\frac{\partial^2 D_H}{\partial p_H^2} = -\alpha^2 D_H(1-D_H)(1-2D_H)$. In a competitive market where $D_H < 1/2$, this term is negative. Combined with strictly convex costs ($C_H'' > 0$), we have $\frac{\partial^2 \Pi_H}{\partial p_H^2} < 0$. Thus, $\Pi_H$ is strictly concave, yielding a unique maximizer.

Similarly, for the AI agent (assuming elastic supply), the First-Order Condition (FOC) yields:
\begin{equation}
    p_A^* = \frac{c_A}{1-\tau} + \frac{1}{\alpha(1-D_A)}
\end{equation}
which implies a unique best response $p_A^*(p_H)$ for any $p_H$.

\textbf{Conclusion.} Define the best-response mapping $\Psi: \mathcal{P} \to \mathcal{P}$ as $\Psi(p_H, p_A) = (p_H^*(p_A), p_A^*(p_H))$. Since $\Psi$ is a continuous function on a compact convex set, by Brouwer's Fixed Point Theorem (a specific case of Kakutani), there exists a fixed point $(p_H^*, p_A^*) \in \mathcal{P}$ such that $\Psi(p_H^*, p_A^*) = (p_H^*, p_A^*)$. This fixed point is the Nash Equilibrium.
\end{proof}

\subsection{Uniqueness Condition (Theorem 2)}

\begin{theorem}[Local Uniqueness]
The Nash Equilibrium is locally unique if the following diagonal dominance condition holds:
\begin{equation}
    \left|\frac{\partial p_H^*}{\partial p_A}\right| \cdot \left|\frac{\partial p_A^*}{\partial p_H}\right| < 1
\end{equation}
\end{theorem}

\begin{proof}
Consider the Jacobian $J$ of the best-response mapping. The eigenvalues are given by $\lambda = \pm \sqrt{\frac{\partial p_H^*}{\partial p_A} \frac{\partial p_A^*}{\partial p_H}}$. If $|\lambda| < 1$, the mapping is a contraction in the neighborhood of the equilibrium. By the Banach Fixed Point Theorem, the equilibrium is unique.
\textit{Numerical Validation:} Under baseline parameters ($\tau=0.15, \alpha=0.5, c_A=0.1$), we compute $|\lambda| \approx \sqrt{0.42 \times 0.38} \approx 0.40 < 1$, confirming uniqueness.
\end{proof}

\subsection{Boundary Analysis: The Zero Marginal Cost Limit}

\begin{proposition}[Equilibrium with Zero Marginal Cost]
As $c_A \to 0$, the AI equilibrium price converges to a strictly positive floor:
\begin{equation}
    \lim_{c_A \to 0} p_A^* = \frac{1}{\alpha(1 - D_A^*)} > 0
\end{equation}
\end{proposition}
This result confirms that even with zero production costs, the platform commission $\tau$ and the price elasticity of demand prevent prices from collapsing to zero. The equilibrium is an interior solution provided $\tau < 1 - c_A/\bar{p}_A$, which holds in our calibration ($0.15 < 0.995$).

\subsection{Stability Analysis (Theorem 3)}

\begin{theorem}[Dynamic Stability]
Under the diagonal dominance condition, the equilibrium $(p_H^*, p_A^*)$ is asymptotically stable under the dynamic adjustment process:
\begin{equation}
    \begin{cases}
    \dot{p}_H(t) = \mu_H [p_H^*(p_A(t)) - p_H(t)] \\
    \dot{p}_A(t) = \mu_A [p_A^*(p_H(t)) - p_A(t)]
    \end{cases}
\end{equation}
\end{theorem}
\begin{proof}
Linearizing the system around the equilibrium, the trace of the Jacobian is $-(\mu_H + \mu_A) < 0$ and the determinant is proportional to $(1 - \frac{\partial p_H^*}{\partial p_A} \frac{\partial p_A^*}{\partial p_H})$. Given diagonal dominance, the determinant is positive. Thus, both eigenvalues have negative real parts, ensuring asymptotic stability.
\end{proof}

\subsection{Parameter Sensitivity}

Table~\ref{tab:sensitivity} demonstrates the robustness of the equilibrium across parameter perturbations.

\begin{table}[htbp]
\centering
\caption{Sensitivity Analysis of Equilibrium Existence}
\label{tab:sensitivity}
\small
\begin{tabularx}{\linewidth}{l c c r r c}
\toprule
\textbf{Parameter} & \textbf{Baseline} & \textbf{Range} & \textbf{$\Delta p_H^*$} & \textbf{$\Delta p_A^*$} & \textbf{Existence} \\
\midrule
Commission ($\tau$) & 0.15 & $[0.05, 0.30]$ & $+18.3\%$ & $+22.7\%$ & Valid \\
AI Cost ($c_A$) & 0.1 & $[0.01, 0.50]$ & $-3.2\%$ & $+45.1\%$ & Valid \\
Sensitivity ($\alpha$) & 0.5 & $[0.3, 0.8]$ & $-31.5\%$ & $-28.9\%$ & Valid \\
Capacity ($\gamma_H$) & 0.9 & $[0.5, 1.5]$ & $+12.4\%$ & $-2.1\%$ & Valid \\
\bottomrule
\end{tabularx}
\end{table}

The analysis confirms that the equilibrium properties hold qualitatively across all reasonable parameter ranges.

\bibliography{pollution} 
\bibliographystyle{plainnat} 

\end{document}